\documentclass[12pt,final]{elsart}

  \usepackage{latexsym,bm,amsmath,amssymb,amsfonts}
  \usepackage{epsfig,graphics,graphicx,subfigure}
  \usepackage{cite}
  \usepackage{slashed}

  \long\def\comment#1{ }
  \newcommand{\dif}{{\rm d}}
  
  \newcommand{\abar}{\bar{\alpha}}

  \newcommand{\mcal}{\mathcal}
  \newcommand{\rme}{{\rm e}}
  \newcommand{\rmi}{{\rm i}}

  \newcommand{\grad}{\nabla}
  \newcommand{\Lam}{\Lambda_{{\rm QCD}}}
  
  \newcommand{\order}[1]{\mcal{O}{(#1)}}
  \newcommand{\nn}{\nonumber\\}

  \newcommand{\beq}{\begin{eqnarray}}
  \newcommand{\eeq}{\end{eqnarray}}

\begin{document}

\vspace*{1.0cm}

\begin{frontmatter}\parbox[]{16.0cm}{\begin{center}
\title{\rm \LARGE Geometric scaling in Mueller-Navelet jets}

\author{E.~Iancu $^{\rm a,1,2}$},
\author{M.S.~Kugeratski $^{\rm a,1}$},
\author{D.N.~Triantafyllopoulos $^{\rm b,1}$}

\address{$^{\rm a}$ Service de Physique Th\'eorique de Saclay,
CEA/DSM/SPhT, F-91191 Gif-sur-Yvette, France}
\address{$^{\rm b}$ ECT$^*$, Villa Tambosi, Strada delle Tabarelle
286, I-38050 Villazzano (TN), Italy}

\thanks{{\it E-mail addresses:}
edmond.iancu@cea.fr (E.~Iancu), maria.kugeratskisouza@cea.fr
(M.S. Kugeratski), dionysis@ect.it (D.N.~Triantafyllopoulos).}
\thanks{Membre du Centre National de la Recherche Scientifique
(CNRS), France.}

\vspace*{1cm}
\begin{abstract}
We argue that the production of Mueller--Navelet jets at the LHC
represents a convenient environment to study gluon saturation and
high--energy scattering in the presence of unitarity corrections. We show
that, in a suitable range of transverse momenta for the produced jets,
the cross--section for the partonic subprocess should exhibit geometric
scaling. We point out that, in the presence of a running coupling, the
cross--section for producing hard jets cannot be fully computed in
perturbation theory, not even after taking into account the saturation
effects: the non--perturbative physics affects the overall normalization
of the cross--section, but not also its geometric scaling behavior.
\end{abstract}
\end{center}}

\end{frontmatter}
\newpage

\section{Introduction}
\setcounter{equation}{0} \label{SECT:INTRO}

Geometric scaling is one of the most important manifestations of the
saturation physics in QCD at high energy
\cite{geometric,SCALING,MT02,MP031}, with striking consequences for the
phenomenology. Identified first \cite{geometric} in the HERA data, via a
phenomenological analysis inspired by the idea of saturation
\cite{GBW99}, geometric scaling has been soon after understood
\cite{SCALING,MT02,MP031} as a property of the BFKL evolution \cite{BFKL}
in the presence of saturation. For problems with a single (transverse)
resolution scale $Q^2$, so like deep inelastic scattering (DIS) or
inclusive single--particle production in hadron--hadron collisions,
`geometric scaling' means that, within a wide kinematical window, the
cross--section scales as a function of the ratio $Q^2/Q^2_s(Y)$. Here,
$Q_s(Y)$ is the {\em saturation momentum}, i.e., the characteristic
momentum scale for the onset of unitarity corrections in a collision in
which the projectile and the target are separated by a rapidity gap $Y$.
This scale grows rapidly with $Y$, according to the BFKL evolution
\cite{SCALING,MT02,DT02}.

Most importantly, geometric scaling is not restricted to the saturation
region at $Q^2\lesssim Q^2_s(Y)$, where the gluon occupation numbers are
large and the scattering amplitudes are close to their unitarity limits,
but it also extends over a relatively wide `geometric scaling window'
\cite{SCALING,MT02} at $Q^2\gg Q^2_s(Y)$, where the target is dilute and
the scattering is weak, yet the scattering amplitudes `feel' the effects
of saturation, via the boundary condition at $Q^2\sim Q^2_s(Y)$. This
scaling window, which with increasing $Y$ is pushed towards larger and
larger values of $Q^2$ (because of the corresponding rise in $Q^2_s(Y)$)
and whose width is slowly increasing with $Y$ (via the BFKL diffusion),
is essentially the same as the validity range for the BFKL approximation
\cite{BFKL} at high energy. Hence, the large--$Q^2$ form of geometric
scaling is a direct consequence of the BFKL dynamics precursory of
saturation \cite{SCALING,MT02,MP031} and can be used to test the latter
at the level of the phenomenology.

So far, the most compelling such tests have been performed on the HERA
data at small values of Bjorken's $x\simeq Q^2/s$, which are the data for
which geometric scaling has been originally identified
\cite{geometric,GS2} (see also Refs. \cite{MS06,GPSS06} for recent
analyses, which include the diffractive data). Namely, one found that,
within the whole small--$x$ domain at HERA, i.e., for $x\le 0.01$ and
$Q^2\le 450$ GeV$^2$, the DIS cross--section can be well approximated by
a scaling function: $\sigma(x,Q^2)\approx \sigma(Q^2/Q^2_s(Y))$ with
$Q^2_s(Y)\sim \rme^{\lambda Y}$ and $Y=\ln(1/x)$. (For comparison, the
proton saturation momentum at HERA is estimated in the ballpark of 1
GeV.) Remarkably, the value $\lambda\simeq 0.3$ for the `saturation
exponent' emerging from these analyses is in rough agreement with its
perturbative calculation \cite{DT02} from the next--to--leading order
BFKL equation \cite{NLBFKL,Salam99}. More detailed analyses
\cite{IIM03,FS04,GS07,Watt07}, combining BFKL dynamics and unitarity
corrections within the framework of the dipole picture, have shown that
the HERA data are consistent with some of the hallmarks of the BFKL
evolution, like its characteristic `anomalous dimension', or the
violation of geometric scaling via BFKL diffusion. Similar
parametrizations for the dipole cross--section, with the parameters fixed
through fits to the HERA data, have been used
\cite{KKT2,Dumitru1,MariaRHIC} to describe particle production in
deuteron--nucleus collisions at RHIC, with some success in explaining the
`high--$p_T$ suppression' in the nuclear modification factor at forward
rapidities.

However, given the kinematical limitations inherent in the experiments at
HERA and RHIC, the previous phenomenological studies of BFKL physics,
geometric scaling, and saturation cannot be viewed as definitive. The
situation should be more favorable in this respect at LHC, where the
higher available energies and the experimental setup should offer larger
rapidity gaps to the BFKL evolution. For instance, in forward particle
production at LHC one could measure values of $x$ as small as $x\sim
10^{-6}$ in the `target' proton wavefunction for a produced jet with
transverse momentum $k_\perp\sim 10$ GeV. With this kinematics, the jet
should explore the geometric scaling window of the target proton, with
interesting consequences, e.g., for the nuclear modification factor
\cite{IIT04,IT07}.

Another interesting process that was proposed to test the BFKL dynamics
and which could be measured at LHC under favorable conditions is the
production of Mueller--Navelet jets \cite{MN87} (see also Refs.
\cite{Duca94,Duca95,Duca95b,Stirling94,Orr97,Andersen01,Andersen03,Vera06,Vera07}
for various theoretical studies and \cite{D0coll96,D0coll00} for
experimental searches at the Tevatron). This is a pair of jets separated
by a large rapidity gap $Y$ which should favor the BFKL evolution of the
cross--section for the partonic subprocess. For sufficiently large values
of $Y$, saturation effects (in the form of unitarity corrections to the
partonic scattering) should become important, as already emphasized by
Mueller and Navelet in their original proposal \cite{MN87}. However, with
the exception of a few, preliminary, phenomenological studies
\cite{MP04,MR06}, such effects have been left out in previous studies of
the Mueller--Navelet jets, which focused on the energies at the Tevatron.
In particular, the modification of the BFKL dynamics by saturation and
the phenomenon of geometric scaling have never been addressed in this
context. These are the aspects that we would like to focus on in what
follows.

Our first observation is that the Mueller--Navelet process is
particularly favorable to study saturation physics. Unlike in DIS, where
the gluon evolution inside the proton starts at the `soft' scale
$\Lam\sim 250$ MeV and thus requires a relatively large rapidity
evolution before it develops a hard saturation momentum, in the context
of Mueller--Navelet jets this evolution starts with the `hard' scale set
by the transverse momentum $k_\perp\ge 10$ GeV of one of the two jets.
Hence, the subsequent evolution with $Y$ should rapidly produce a system
with very high gluon density around the position of the jet, i.e., with a
very large {\em local} saturation momentum. The counterpart of that is
that the dense region occupies only a small area $\sim 1/k_\perp^2$ in
impact parameter space, and looks like a `dense spot'.

To be more specific, recall that one needs a rapidity evolution
$Y_0\simeq (1/\omega_\mathbb{P})\ln(1/\alpha_s^2)$, with
$\omega_\mathbb{P}$ the BFKL intercept, before a small hadronic system,
so like a dipole or a high--momentum parton, reaches saturation on the
resolution scale set by its own size, or transverse momentum \cite{AM95}.
A leading--order estimate for $\omega_\mathbb{P}$ would yield $Y_0\simeq
5$ (for $\alpha_s=0.2$, as appropriate for a 10 GeV jet), but this is
probably too optimistic. A more realistic estimate, using the NLO BFKL
intercept \cite{Salam99}, is $Y_0\simeq 8$, which is fully within the
reach of LHC. This means that, when producing a pair of Mueller--Navelet
jets separated by a rapidity gap $Y=8$ at the LHC, then one of the jets
will `see' the other one as a high--density gluonic system (a `color
glass condensate') with a saturation momentum $Q_s(Y)\sim 10$ GeV.
Moreover, every additional unit of rapidity will make this saturation
scale even harder, according to $Q_s^2(Y)\simeq
Q_0^2\exp[\lambda(Y-Y_0)]$. Such values for $Q_s$ are considerably higher
than those that could ever be achieved in a proton, or nuclear,
wavefunction at LHC energies, even for the most forward collisions.

Furthermore, the evolution towards saturation should favor the
transverse momentum dissymmetry between the two jets\footnote{We recall that in lowest--order perturbation theory the two jets
come out with equal and opposite transverse momenta, because of momentum
conservation. Hence, any momentum asymmetry between the jets is a signal
of a multiparticle final state, as produced in particular by the BFKL evolution.}. The typical transverse momentum of a gluon within the
wavefunction of an evolved jet is the respective saturation momentum
$Q_s(Y)$, which for large $Y$ is as hard as, or even harder than, the
original parton which comes out as a jet. Therefore, the other jet can
easily be produced with a very different transverse momentum, because the
momentum imbalance can be compensated by `inclusive' gluons from the
`evolved' jet wavefunction. In fact, we expect a larger momentum
asymmetry in the presence of saturation than from pure BFKL evolution:
unlike the latter, which proceeds symmetrically towards soft and hard
momenta, the evolution in the presence of saturation is biased towards
large transverse momenta (larger than the saturation scale).

The most interesting kinematical situation for our subsequent analysis is
precisely when the two jets are well separated in transverse momentum,
say, $k_{1\perp}\gg k_{2\perp}$. More precisely, we shall be interested
in configurations where the harder jet has a transverse momentum
comparable to, or even larger than, the saturation momentum that would be
generated by the evolution of the softer jet over the rapidity gap $Y$:
$k_{1\perp}\gtrsim Q_s(Y)$, where $Q_s(Y)$ is implicitly a function of
$k_{2\perp}$. Under these circumstances, we shall see that the partonic
cross--section exhibits geometric scaling within a wide kinematical
window. That is, for given $k_{2\perp}$, the cross--section scales as a
function of the ratio $k_{1\perp}^2/Q_s^2(Y)$. On the other hand, the
cross--section is small, of order $1/k_{2\perp}^2$, because of the small
size of the dense spot, as alluded to above. While our conclusions may
look natural, given the kinematics and the similarity with other problems
like DIS, our analysis appears to reserve some difficulties and
surprises.

The first difficulty refers to the inclusion of saturation effects and
unitarity corrections in the cross--section for Mueller--Navelet jets.
This in turn requires two steps: \texttt{(i)} a factorization formula
which is general enough to allow for unitarity corrections, and
\texttt{(ii)} the calculation of the ingredients which enter this
factorization formula within the framework of high--density QCD (i.e.,
from the solutions to the non--linear evolution equations which
generalize the BFKL equation to the region of high gluon density).

Concerning step \texttt{(i)}, we shall proceed in a heuristic way, by
generalizing, in Sect. 2, a known formula for single--jet production in
the presence of unitarity corrections \cite{KM98,KTS99,KT02,BGV04,CM04}.
This leads us to a generalization of the standard $k_T$--factorization
for Mueller--Navelet jets, which has been already presented in Ref.
\cite{CM04,MP04,MR06}, and in which the BFKL Green's function is replaced
by the total cross--section for the scattering between two {\em
effective} color dipoles. Unlike the quark--antiquark dipole familiar in
the context of DIS (see, e.g., \cite{GBW99,IIM03,FS04,GS07,Watt07}),
which is a physical fluctuation of the virtual photon, the dipoles that
enter our factorization for Mueller--Navelet jets are merely mathematical
constructions, which appear in the calculation of the cross--section and
are built with one parton in the direct amplitude and another parton in
the complex conjugate amplitude. The dipole--dipole cross--section is
written in coordinate space, as appropriate for the inclusion of
unitarity corrections (multiple scattering) in the eikonal approximation.
The transverse momenta $k_{1\perp}$ and $k_{2\perp}$ of the produced jets
are then fixed via a double Fourier transform from the dipole sizes.

The second step, i.e., the calculation of the dipole--dipole
cross--section within high--density QCD, turns out to be particularly
subtle. Since the scattering involves two systems (dipoles) which start
by being dilute at low energy, it seems that we cannot rely on the
standard Balitsky--JIMWLK, or BK, equations \cite{B,K,JKLW,CGC}, which
apply only to dense--dilute scattering. Instead, one should use the more
general, `Pomeron loop', equations \cite{IT04,MSW05,KL05a}, which also allow for particle number fluctuations in the course of the evolution. From the correspondence with statistical physics \cite{IMM04}, and also from the numerical simulations of simple models inspired by QCD
\cite{GS05,ISST07}, we know that, with a {\em fixed} coupling, the
effects of the fluctuations are truly crucial: with increasing energy,
they rapidly wash out both the BFKL approximation and the `geometric
scaling' behavior predicted by the BK equation
\cite{SCALING,MT02,MP031}. However, a very recent numerical analysis
\cite{RPLOOP} has shown that the fluctuations are strongly suppressed by
the {\em running} of the coupling, in such a way that their effects
remain negligible for all energies of practical interest.

With the philosophy that the running--coupling case is the only one of
fundamental interest for real QCD, in what follows we shall perform a
`mean field' type of analysis, based on BK equation, for both fixed and
running coupling. The fixed--coupling analysis, as developed in Sects. 3
and 4, turns out to be rather straightforward: The dipole--dipole
cross--section, as obtained from approximate solutions to BK equation
\cite{SCALING,MT02,MP031}, exhibits geometric scaling for suitably chosen
dipole sizes. After a Fourier transform, this scaling property gets
transmitted to the partonic core of the Mueller--Navelet cross--section,
for appropriate transverse momenta of the two jets.

The running--coupling case, that we shall treat in Sect. 5, is both more
interesting and more subtle. First, it might look inconsistent to include
the running of the coupling, but at the same time ignore other
next--to--leading order corrections in perturbative QCD. But it turns
that the running of the coupling plays a special role in the context of
the high--energy dynamics: because of it, all the other perturbative
corrections die away in the high--energy limit \cite{DT02,BP07}. Indeed,
the evolution towards saturation is controlled by momenta $k_\perp\sim
Q_s(Y)$; so, the relevant value of the coupling is $\alpha_s(Q_s^2(Y))$,
which decreases with $Y$, and therefore so do the perturbative
corrections\footnote{Of course, these corrections may be numerically
important for the phenomenology at LHC, but here we focus on the dominant
asymptotic behavior, for simplicity.}, whose strength is proportional to
$\alpha_s(Q_s^2(Y))$. Moreover, the running of $\alpha_s$ has qualitative
consequences which modify the high--energy evolution in depth. We have
already mentioned its role in suppressing the particle--number
fluctuations. This is related to a more general property of the running
of the coupling, which is to {\em slow down} the evolution towards
saturation \cite{SCALING,MT02,DT02}. Another manifestation of this
property is in the growth of the saturation momentum with $Y$: for
sufficiently large $Y$, and with a running coupling, $\ln Q_s^2$ grows
like $\sqrt{Y}$, and not like $Y$.

An important consequence of the running of the coupling, which is not
specific to the high--energy problem, but has dramatic consequences for
it, is the fact that it introduces an intrinsic scale in the problem ---
the `soft' scale $\Lam$ ---, thus breaking down the conformal invariance
of the leading--order formalism. With increasing energy, this scale
progressively replaces within the saturation momentum any other scale
introduced by the initial conditions at low energy, so like the target
dipole size. Accordingly, for sufficiently high energy, the saturation
momentum becomes independent of the target size \cite{AM03}. This has
important consequences for the high--energy evolution in general (e.g.,
it implies that a large nucleus is not more dense than a proton at very
high energies), and for the Mueller--Navelet process in particular: it
implies that the perturbative calculation of the Mueller--Navelet
cross--section breaks down, even if the jet transverse momenta are
restricted to be hard. The precise argument in that sense will be
developed in Sect. 4, but here we would like to emphasize that this
argument is in fact very general (and hence also very robust): it
reflects the fact that, with running coupling, the dipole--dipole
scattering amplitude {\em at a fixed impact parameter} is independent of
the target dipole size $R$ (rather than dying away as an inverse power of
$R$, as it would happen in the fixed--coupling formalism, by conformal
invariance). Accordingly, the dipole--dipole cross--section, which is
obtained by integrating the amplitude over all impact parameters, is
proportional to $R^2$, and hence it strongly favors large dipole
fluctuations. Without the non--perturbative cutoff introduced by
confinement, the partonic cross--section would be controlled by dipole
fluctuations with arbitrarily large size. As we shall argue in Sect. 4,
the ad--hoc introduction of a non--perturbative cutoff on the dipole
sizes affects the normalization of the total cross--section, but not also
its property of geometric scaling (which merely refers to the functional
dependencies of the cross--section upon the rapidity gap $Y$ and upon the
transverse momentum $k_{1\perp}$ of the hardest jet).

\section{Forward jets with unitarity corrections}
\setcounter{equation}{0} \label{SECT:JETS}

Although our main interest here is in the production of a {\em pair} of
(Mueller--Navelet) jets, it is instructive to start our presentation with
the case of a single jet, for which the high--energy factorization in the
presence of unitarity corrections is more firmly established. This will
also allow us to introduce the physics and the theoretical description of
the unitarity corrections in a simpler setting. The two--jet problem will
then be easier to explain, by analogy.

\begin{figure}[t]
\begin{center}
\includegraphics[width=7.cm]{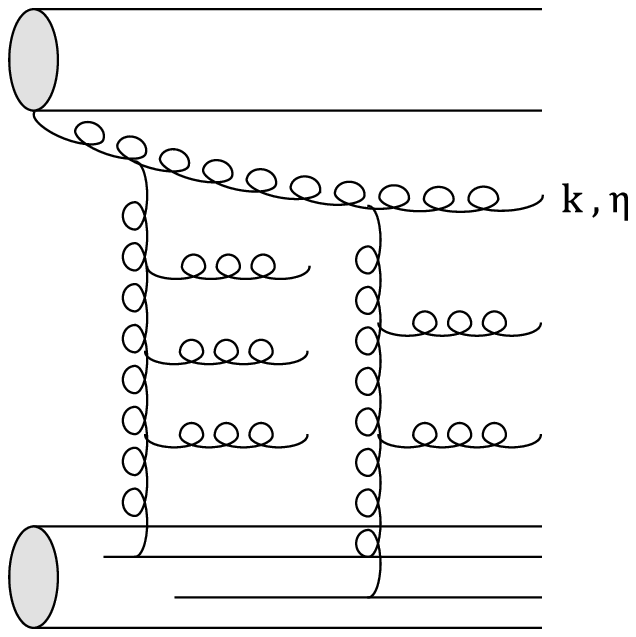}\qquad\quad
\includegraphics[width=7.cm]{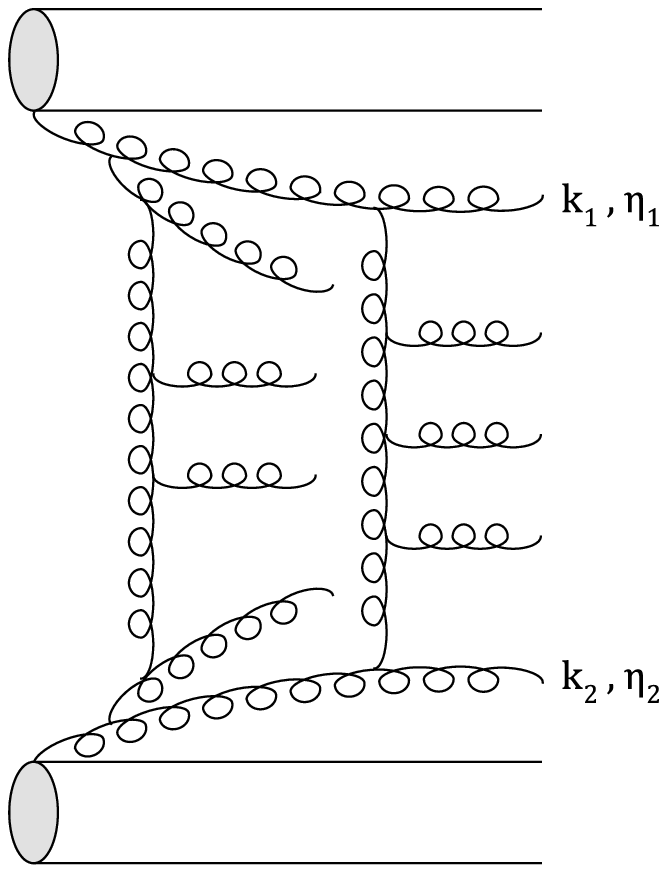}
\vspace*{.3cm} \hspace*{-1.5cm} (a)\hspace*{7.5cm}(b) \caption{\sl The
inclusive production of a single, forward, jet (left) and of a pair of
Mueller--Navelet jets (right) in the presence of unitarity corrections. }
 \label{jets}
\end{center}
\end{figure}

For definiteness, we focus on jets initiated by gluons (quarks will be
added later on), and thus consider the cross--section for inclusive gluon
production at forward rapidity in a hadron--hadron collision at high
energy. By `forward rapidity' we mean that the produced gluon carries a
sizeable fraction $x\sim\order{1}$ of the longitudinal momentum of one of
the incoming hadrons (the `projectile'), so that there is a large
rapidity gap $Y=Y_0-y$ between this produced gluon and the other hadron
(the `target'). Here, $Y_0=\ln(s/M_1M_2)$, with $s$ the invariant energy
squared and $M_{1,2}$ the masses of the participating hadrons, is the
rapidity gap between the projectile and the target, and
$y=\ln(1/x)+\ln(k_\perp/M_1)$ is the (relatively small) rapidity
separation between the produced gluon, which has transverse momentum
$k_\perp$, and the projectile. Alternatively, one could trade the
rapidity gap $y$ for the pseudo--rapidity $\eta$ of the produced jet in
the laboratory frame; e.g., if the lab frame coincides with the hadron
center--of--mass frame, so like at LHC, then
$x=(k_{\perp}/\sqrt{s})\,\rme^{\eta}$, with $\eta> 0$ for forward jets
(see also Fig. \ref{jets}.a).

Under these circumstances, and in the leading--order formalism of
perturbative QCD (meaning, in particular, that the coupling is fixed),
the cross--section for gluon production can be expressed in a
`$k_T$--factorized' form, which is formally similar to, but more general
than, the corresponding factorization used in the context of the BFKL
physics \cite{BFKL}. Namely, the $k_T$--factorization is now extended
towards the high--energy regime where unitarity corrections (multiple
scattering, gluon saturation) become important. The general respective
formula can be found in the literature (see, e.g., Refs.
\cite{KM98,KTS99,KT02,BGV04,CM04,GLUON}). Here, we shall need only a
simpler form of it, valid when the transverse momentum $k_\perp$ of the
produced gluon is large enough --- much larger than the typical momentum
transferred from the projectile to this gluon (see below for a more
precise condition). We then have
 \beq\label{onejet}
    \frac{\dif \sigma^{PT \rightarrow J X}}{\dif \eta
    \,\dif^2k_\perp } \,=\,
    \frac{1}{8\pi^2k_\perp^2}\,xG_P(x,k_\perp^2)
 \int {\dif^2 {\bm r}}\ {\rm
   e}^{-i {\bm k} \cdot {\bm r}}\ \grad^2_r \sigma_{(gg)T}({\bm r},Y)\,,
    \eeq
where $xG_P(x,k_\perp^2)$ is the gluon distribution in the projectile
($P$) on the resolution scale of the jet (i.e., the number of gluons with
longitudinal momentum fraction $x$ equal to that of the jet, and with
transverse momenta $p_\perp^2\le k_\perp^2$). Furthermore,
$\sigma_{(gg)T}({\bm r},Y)$ is the total cross--section for the
scattering between a {\em gluonic} dipole (a $gg$ pair in a color singlet
state) with transverse size ${\bm r}$ and the hadronic target ($T$), for
a rapidity separation $Y$. The $gg$ dipole here is the effective dipole
made with the produced gluon in the direct amplitude (located at
transverse coordinate ${\bm x}$) and the corresponding gluon in the
complex conjugate amplitude (located at ${\bm y}$). The gluon transverse
momentum $k_\perp=|{\bm k}|$ in the final state is then fixed via the
Fourier transform from $\bm{r} =\bm{x}-\bm{y}$ to ${\bm k}$.

In the single--scattering approximation to the dipole--target
cross--section, the Fourier transform in Eq.~(\ref{onejet}) yields the
usual `unintegrated' gluon distribution in the target wavefunction
(evaluated in the BFKL approximation) times a constant of order
$\alpha_s$. We then recover from Eq.~(\ref{onejet}) the traditional
$k_T$--factorization. But Eq.~(\ref{onejet}) remains valid also very
large values of $Y$, where the unitarity corrections to the dipole
scattering become important and the BFKL approximation ceases to apply.

The physical interpretation of the unitarity corrections is most
transparent in the `target infinite momentum frame', where the dipole has
relatively low energy while the target carries most of the total rapidity
$Y$. Then the target wavefunction has evolved into a `color glass
condensate' (CGC) --- a system with high gluon density characterized by a
hard intrinsic scale, the {\em saturation momentum} $Q_s(Y)$, which grows
rapidly with $Y$ and separates between

\vspace*{.2cm}
\begin{itemize}

\item  a high density region at low momenta $p_\perp\lesssim Q_s(Y)$,
    where the gluon occupation numbers are large,
    $\sim\order{1/\alpha_s}$, but `saturated' (they do not rise with
    the energy anymore), and

\item a low density region at high momenta $p_\perp \gtrsim Q_s(Y)$,
    where the occupation numbers are low, but rapidly growing with
    $Y$, via the BFKL evolution.

\end{itemize}

\vspace*{.2cm} \noindent For sufficiently high energy and/or large dipole
sizes, such that $r\gtrsim 1/Q_s(Y)$, the dipole will undergo multiple
scattering off the CGC.  The `unitarity corrections' refer to this
multiple scattering, as well as to the saturation effects in the target
gluon distribution, i.e., to all the {\em non--linear} effects which
reduce the gluon density and enforce the unitarity bound on the
scattering process. Such effects are resummed --- within the eikonal
approximation, and within the limits of the LO formalism --- by the
Balitsky--JIMWLK equations, which in particular determine the dipole
scattering amplitude in this high energy regime\footnote{As mentioned in
the Introduction, these equations neglect the particle number
fluctuations, which would be important in the context of the
fixed--coupling evolution \cite{IMM04,IT04}, but which are suppressed by
the running of the coupling \cite{RPLOOP}.}. The corresponding solution
will be further described in the next sections. Here, it suffices to
notice that the factorization (\ref{onejet}) is correct when $Y$ is so
large that $Q_s^2(Y)\gg \Lam^2$ (in order for the perturbative approach
to unitarity corrections to be valid), and for relatively `hard' jets,
whose transverse momenta are not much smaller than $Q_s(Y)$ --- the
precise condition being $k_\perp^2\gg Q_s(Y)\Lam$. Indeed, under these
circumstances, the contribution shown in Eq.~(\ref{onejet}) is enhanced
with respect to the other, missing, contributions by the large logarithm
$\ln(k_\perp^2/\Lam^2)$ (via the gluon distribution $xG_P(x,k_\perp^2)$).

We are now prepared to present the corresponding formul\ae{} for the
Mueller--Navelet jets. This is a pair of jets produced in a high--energy
hadron--hadron collision such that each jet carries a relatively large
fraction $x_i$, $i=1,2$ of the longitudinal momentum of its parent
hadron. Accordingly, each jet is relatively close in rapidity to its
respective parent, so there is a large rapidity gap $Y\gg 1$ between the
jets: $Y=\ln(x_1x_2s/k_{1\perp}k_{2\perp})$, where $k_{1\perp}$ and
$k_{2\perp}$ are the jet transverse momenta  (see Fig. \ref{jets}.b).
When the scattering is viewed in the hadron center--of--mass frame, the
momentum fractions $x_i$ and the rapidity gap are determined by the
pseudo--rapidities $\eta_i$ of the two jets, according to
 \beq\label{kinem}
 x_1\simeq\,\frac{k_{1\perp}}{\sqrt{s}}\,\rme^{\eta_1}\,,\qquad
 x_2\simeq\,\frac{k_{2\perp}}{\sqrt{s}}\,\rme^{-\eta_2}\,,\qquad
 Y=\eta_1-\eta_2\,.\eeq
The typical kinematics for Mueller--Navelet jets is such that $\eta_1$ is
large and positive, while $\eta_2$ is large and negative.

The differential cross--section for Mueller--Navelet jets has been
rigorously computed \cite{MN87} (within the LO formalism, once again)
only at the level of the BFKL approximation, which ignores unitarity
corrections. The corresponding result is the expected generalization of
the corresponding single--jet cross--section
--- the BFKL version of Eq.~(\ref{onejet}) --- which is symmetric w.r.t.
the two jets. In view of this, and of the symmetry of the problem, it has
been conjectured  \cite{CM04,MP04,MR06} that, after including the
unitarity corrections, the cross--section for  Mueller--Navelet jets
should be given by the properly symmetrized version of
Eq.~(\ref{onejet}), that is
 \beq
  \frac{\dif \sigma^{pp \rightarrow J X J}}
  {\dif x_1 \dif x_ 2 \,\dif^2 k_{1\perp}
  \dif^2 k_{2\perp}} = & & \frac{1}{64\pi^4}
 \,G(x_1,k_{1\perp}^2) \,G(x_2,k_{2\perp}^2)\,
 \frac{1}{k_{1\perp}^2 \, k_{2\perp}^2} \nn & & \times
 \int {\dif^2 {\bm r}_1}\, {\rm
   e}^{-i {\bm k}_1 \cdot {\bm r}_1}\!\! \int {\dif^2 {\bm r}_2}\, {\rm
   e}^{-i {\bm k}_2 \cdot {\bm r}_2} \,
\nabla_{r_1} ^2  \nabla_{r_2} ^2 \sigma_{(gg)(gg)}  ( {\bm r}_1, {\bm
r}_2, Y), \label{MVjets}
\eeq
where, for definiteness, we have chosen the incoming hadrons to be
protons, so like at LHC. $\sigma_{(gg)(gg)}({\bm r}_1,  {\bm r}_2, Y)$ is
the total cross--section for the scattering between two (effective)
gluonic dipoles with the indicated transverse sizes and separated by a
rapidity gap $Y$.

The cross--section \eqref{MVjets} is generally a function of
$k_{1\perp}$, $k_{2\perp}$, and the relative angle $\phi$ between the
vectors ${\bm k}_1$ and ${\bm k}_2$ (the azimuthal angle between the two
jets). Although interesting in view of the phenomenology (and largely
studied in the context of the BFKL approximation
\cite{Duca94,Duca95,Duca95b,Stirling94,Orr97,Andersen01,Andersen03,Vera06,Vera07}),
the azimuthal correlations represent a subleading effect at high
energies, and thus are irrelevant for our present study of the unitarity
corrections. So, in what follows we shall average over $\phi$, which is
tantamount to replacing the dipole cross--section in \eqref{MVjets} by
$\sigma_{(gg)(gg)}(r_1, r_2, Y)$ --- the corresponding cross--section
averaged over the relative orientations of the two dipoles.

It is furthermore convenient to introduce some hard momentum cutoffs
$Q_1$ and $Q_2$ (corresponding to the experimental $k_T$--cuts) and
compute the cross--section for producing two jets with transverse momenta
$k_{1\perp}>Q_1$ and $k_{2\perp}>Q_2$ and with given longitudinal
momentum fractions $x_1$ and $x_2$. That is,
   \beq\label{j1j1def}
    \frac{\dif \sigma^{pp \rightarrow J X J}}{\dif x_1 \dif x_2}
     \,\equiv \int {\dif^2 {\bm k}_1}\int {\dif^2 {\bm k}_2}\,
     \frac{\dif \sigma^{pp \rightarrow J X J}}
  {\dif x_1 \dif x_ 2 \,\dif^2 k_{1\perp}\dif^2 k_{2\perp}}\
  \Theta(k_{1\perp}-Q_1)\,\Theta(k_{2\perp}-Q_2)\,.\eeq
(These integrations automatically implement the average over $\phi$.) A
priori, the integrations are complicated by the $k_\perp$--dependencies
of the gluon distributions in \eqref{MVjets}, and by the one implicit in
the dipole--dipole cross--section, via the rapidity gap
$Y=\ln(x_1x_2s/k_{1\perp}k_{2\perp})$. Note however that for given
$x_1,\,x_2$ and $s$, the maximal rapidity gap $Y_{\rm
max}=\ln(x_1x_2s/Q_1Q_2)$ is attained for the threshold momenta
$k_{1\perp}=Q_1$ and $k_{2\perp}=Q_2$. Hence the dominant contribution,
in the sense of the leading--logarithmic approximation, is obtained by
replacing $Y\to Y_{\rm max}$ within $\sigma_{(gg)(gg)}$. Moreover, the
integrand in \eqref{j1j1def} is rapidly decreasing at very large values
for $k_{1\perp}$ and $k_{2\perp}$ --- this can be checked, e.g., by using
the BFKL approximation for the dipole--dipole cross--section \cite{MN87}
---, which enables us to replace $G(x_i,k_{i\perp}^2)\to G(x_i,Q_i^2)$ in
the slowly varying gluon distributions. The remaining integrations can be
easily performed, with the final result
 \beq\label{j1j1}
    \frac{\dif \sigma^{pp \rightarrow J X J}}
    {\dif x_1 \dif x_2}\,\simeq
 \,& & \frac{1}{64\pi^4}\,G(x_1,Q_1^2) \,G(x_2,Q_2^2)\nn
  & & \qquad\times  \int\limits_0 ^{\infty}
    \dif r_1 \int\limits_0 ^{\infty} \dif r_2\,
    Q_1 J_1 (Q_1 r_1) Q_2 J_1 (Q_2 r_2) \sigma_{(gg)(gg)}(r_1, r_2, Y).
    \eeq
Here and from now on it is understood that $Y=\ln(x_1x_2s/Q_1Q_2)$. So
far, we have considered only gluon jets, but quarks or antiquarks jets
can be similarly included: when the jet $i$, with $i=1,2$, is initiated
by a quark with flavor $f$, we have a formula similar to Eq.~(\ref{j1j1})
in which the gluon distribution $G(x_i,Q_i^2)$ is replaced by the quark
distribution $q_f(x_i,Q_i^2)$ (or $\bar q_f(x_i,Q_i^2)$ for an
antiquark), and the corresponding dipole within $\sigma_{(gg)(gg)}$ is
replaced by a dipole made with a quark and an antiquark. We thus
encounter three types of dipole--dipole processes: $(gg)(gg)$,
$(gg)(q\bar q)$, and $(q\bar q)(q\bar q)$, whose cross--sections differ
at most through color factors (see below).

As a consistency check of our above factorization of the Mueller--Navelet
cross--section, cf. Eq.~\eqref{MVjets} or (\ref{j1j1}), let us now verify
its BFKL limit. For two gluonic dipoles, the BFKL cross--section
(averaged over angle) reads\footnote{The appearance of the Casimir $C_F$
for the fundamental representation in a cross--section pertinent to
gluons alone may look surprising. In reality, this has been generated via
the identity $N_c^2/(N_c^2-1)=N_c/2C_F$, where all the $N_c$ factors
arise from the gluon color algebra.}
   \beq\label{sbfkl}
    \sigma_{(gg)(gg)} (r_1, r_2, Y)\Big|_{\rm BFKL}\, =
     {2 \pi \alpha_s ^2}\,\frac{N_c}{C_F}\,
     r_1 ^2 \int \frac{\dif \gamma}{2 \pi \rmi}\,
    \frac{(r_2 / r_1)^{2 \gamma}}{\gamma^2 (1 - \gamma)^2}
    \exp{\left[ \frac{\alpha_s N_c}{\pi} \chi (\gamma) Y \right]},
    \eeq
where we use the standard representation for the BFKL solution in Mellin
space (see, e.g., \cite{SCALING,MT02}). This result is symmetric under
the exchange $r_1\leftrightarrow r_2$ of the two dipoles, as it can be
checked by using the property $\chi(\gamma)=\chi(1-\gamma)$ of the BFKL
characteristic function. When one or both of the gluonic dipoles are
replaced by fermionic ($q\bar q$) ones, the expression in
Eq.~(\ref{sbfkl}) must be multiplied by a factor ${C_F}/{N_c}$ for each
such a replacement.

Substituting \eqref{sbfkl} into \eqref{j1j1}, making use of the Bessel
function integration formula
    \beq\label{jpower}
    \int\limits_0 ^{\infty} \dif x\, x^\beta J_n(x) = 2^\beta\,
    \frac{\Gamma \Big(\displaystyle{\frac{n+1+\beta}{2}}\Big)}
    {\Gamma \Big(\displaystyle{\frac{n+1-\beta}{2}}\Big)},
    \eeq
and summing up over all types of partons, we obtain (with $\abar\equiv
{\alpha_s N_c}/{\pi}$)
    \beq\label{MNJBFKL}
    \frac{\dif \sigma^{pp \rightarrow J X J}}
    {\dif x_1 \dif x_2}\bigg|_{\rm BFKL}\,
    = F_{\rm eff}\,\,\frac{{8 \pi}N_c}{C_F}\,
    \frac{\alpha_s ^2}{Q_1 ^2}\,
    \int \frac{\dif \gamma}{2 \pi \rmi}\,
    \frac{(Q_1 ^2 / Q_2 ^2)^{\gamma}}{\gamma (1 - \gamma)}
    \exp{\left[ \abar \chi (\gamma) Y \right]},
    \eeq
where $F_{\rm eff}$ involves contributions from quarks, antiquarks, and
gluons, with appropriate color factors (the sum over the quark flavors is
kept implicit):
  \beq\label{Feff}
  F_{\rm eff}&\,=\,&\frac{1}{64\pi^4}\,f_{\rm eff}(x_1,Q_1^2)\,
  f_{\rm eff}(x_2,Q_2^2)\nn
  f_{\rm eff}(x,Q^2)&\,\equiv\,&G(x,Q^2)\,+\,
  \frac{C_F}{N_c}\big[q(x,Q^2)+\bar q(x,Q^2)\big]
  \,.\eeq
As anticipated, Eq.~\eqref{MNJBFKL} is in precise agreement (including
the normalization) with the corresponding result in Ref. \cite{MN87}.

\section{Fixed coupling case: the dipolar cross--section}
\label{Sect:FCDIP} \setcounter{equation}{0}

In addition to taming the BFKL growth of the dipole scattering amplitude,
in compliance with the unitarity bound, the non--linear effects encoded
in the Balitsky--JIMWLK (or BK) equations have also an interesting
consequence for the functional form of the amplitude in the transition
region from weak to strong scattering: within a rather wide kinematical
region, whose width is increasing with $Y$, this amplitude shows {\em
geometric scaling} \cite{SCALING,MT02}, i.e., it depends upon the size
$r$ of the projectile dipole and upon $Y$ only via the dimensionless
variable $\tau\equiv r^2Q_s^2(Y)$, with $Q_s(Y)$ the saturation momentum
of the target evolved up to rapidity $Y$. It is then tempting to
conjecture that this scaling property should transmit from the
dipole--dipole amplitude to the cross--section for Mueller--Navelet jets,
via the convolutions in Eq.~(\ref{j1j1}) (within a suitabe range of
values for the variables $Q_1$, $Q_2$, and $Y$). This is a simple
argument in that sense: due to the presence of the rapidly oscillating
Bessel functions in the integrand of Eq.~(\ref{j1j1}), one expects the
integrals there to be dominated by values $r_1\sim 1/Q_1$ and $r_2\sim
1/Q_2$. If this is true, then one can choose $Q_1$ and $Q_2$ (for a given
$Y$) in such a way that the dipole--dipole cross--section, and hence the
dijet cross--section, are in the geometric scaling window. As we shall
later discover, via explicit calculations, this simple argument is indeed
correct in the case of a fixed coupling, but not also for a running
coupling.

We start with the fixed--coupling case, i.e., the LO formalism. Within
the eikonal approximation, the dipole--dipole cross--section is computed
as (for dipoles made with partons in a generic color representation)
 \beq\label{sigmagg}
  \sigma_{\rm dd}\,(r_1,r_2,Y)\,=\,2\int\dif^2{\bm b}\ T_{\rm dd}
 ({\bm r}_1,{\bm r}_2,{\bm b},Y)\,,\eeq
where $T_{\rm dd}({\bm r}_1,{\bm r}_2,{\bm b},Y)$ is the scattering
amplitude for two dipoles with transverse sizes ${\bm r}_1$ and ${\bm
r}_2$, relative impact parameter ${\bm b}$, and rapidity separation $Y$.
(The average over the relative angle between ${\bm r}_1$ and ${\bm r}_2$
is implicit here and from now on.) We use conventions in which the
$S$--matrix is written as $S=1-T$, where $T$ is taken to be real, as
appropriate for the dominant behavior at high energy. In general, the
dipole--dipole amplitude and cross--section are, of course, symmetric
under the exchange of the two dipoles, but their approximate forms that
we shall derive below are valid only when one of the dipoles is much
smaller than the other one. It is then convenient to introduce the
notations $r \equiv\min(r_1,r_2)$ and $R \equiv\max(r_1,r_2)$, and refer
to the small (large) dipole as the `projectile' (respectively, the
`target'). Also, as explained in Sect. \ref{SECT:JETS}, it is convenient
to visualize the evolution with increasing energy as gluon evolution in
the `target' (the larger dipole).

The unitarity of the $S$--matrix implies $T\le 1$, with the upper bound
$T=1$ (the `black disk limit') describing a situation where the
scattering occurs with probability one. This constraint is indeed obeyed
by the solution $T$ to the Balitsky--JIMWLK (or BK) equations, which,
moreover, appears to saturate the black disk limit $T=1$ for sufficiently
large $Y$.  But the cross--section \eqref{sigmagg} can rise indefinitely
with $Y$, even after the `black disk' limit has been reached at central
impact parameters, because the gluon distribution in the target keeps
expanding towards larger impact parameters, due to the non--locality of
the BFKL evolution. This radial expansion of the gluon distribution is
however much slower than its evolution towards the black disk limit at a
fixed value of ${\bm b}$. In particular, we do not expect this expansion
to be essential at the LHC energies. Therefore, a target which at $Y=0$
starts as a single dipole of size $R$, evolves with $Y$ towards
blackness, first, on its own scale $R$, then, on smaller and smaller
scales, without significantly expanding towards larger
sizes\footnote{This property is not correctly encoded in the
Balitsky--JIMWLK, or BK, equations, which rather predict a rapid radial
expansion of the black disk, because of the long--range tails (in $b$) of
the perturbative gluon distribution \cite{KW02}. However, in real QCD we
expect this tails to be cut--off by confinement, with the effect that the
radial expansion is drastically slowed down, in compliance with Froissart
bound \cite{FB}.}. Accordingly, when this evolved target is probed by a
projectile dipole with size $r\ll R$, the amplitude $T_{\rm dd}(r,R,b,Y)$
is negligibly small when the two dipoles have no overlap with each other
($b\gg R$). The typical impact parameters which contribute to the
cross--section are such that $b\ll R$, and for them the amplitude is
roughly independent of $b$. These considerations motivate the following
approximation to the dipole--dipole cross--section \eqref{sigmagg}, valid
when $r\ll R$
 \beq\label{Tgg}
  \sigma_{\rm dd}\,(r,R,Y)\,\simeq\,2\pi R^2\, T_{\rm dd}
 (r,R,Y)\,,\eeq
where $T_{\rm dd}(r,R,Y)$ is independent of $b$ and satisfies the
unitarity bound $T_{\rm dd}\le 1$. For this amplitude, we shall use
approximate solutions to the BK equation with fixed coupling.

Specifically, the saturation momentum $Q_s(R,Y)$ is defined by the
condition
 \beq\label{Tsat}
T_{\rm dd}(r,R,Y)\,=\,\kappa\qquad{\rm for}\qquad
 r\,=\,1/Q_s(R,Y)\,,\eeq
where $\kappa < 1$ is a number of order one (its precise value is
irrelevant to the accuracy of interest). For $r\gtrsim 1/Q_s$ we have
$T_{\rm dd}\sim\order{1}$, whereas for $r\ll 1/Q_s$ the amplitude is
small, $T_{\rm dd}\ll 1$, and approximately given by the following, {\em
universal}, function
 \beq\label{TSAT}
T_{\rm dd}(r,R,Y)\,\simeq\,\,\left(\ln\frac{1}{r^2 Q_s^2}\right)\,
 ({r^2}{Q_s^{2}})^{\gamma_s}\,
 \exp\left\{-\frac{\ln^2(r^2 Q_s^2)}
 {4D_s\abar Y} \right\}\eeq
(up to a normalization factor), with the saturation momentum
  \beq\label{QSAT}
  Q_s^2(R,Y)\,=\,(C\alpha_s^2)^{1/\gamma_s}
  \,\frac{1}{R^2}\,\rme^{\lambda_s\abar Y}\,.
  \eeq
The various exponents which appear in these formul\ae{} are pure numbers
determined by the BFKL characteristic function (see Refs.
\cite{SCALING,MT02} for details). Specifically, $\gamma_s\approx 0.63$
($1-\gamma_s\approx 0.37$ is the BFKL anomalous dimension at saturation),
$D_s\approx 48.5$ plays the role of a diffusion coefficient, and
$\lambda_s\approx 4.88$ is the saturation exponent. Note that rapid
growth of the saturation momentum with $Y$. Its dependence upon the
target size $R$ could have been anticipated from dimensional arguments.

Eq.~\eqref{TSAT} is universal in the sense that the dependence upon the
initial conditions at low energy is fully encoded in the value of the
saturation momentum. But the latter is, of course, process--dependent: it
depends upon the size $R$ of the target, and also upon the color
representations of the partons making up the two dipoles participating in
the collision (via the coefficient $C$ which here is left unspecified).
For instance, the value of $C$ corresponding to the process $(gg)(q\bar
q)$ is larger by a factor $N_c/C_F$ than that for the process $(q\bar
q)(q\bar q)$.

The prefactor involving $\alpha_s^2$ in the expression \eqref{QSAT} for
the saturation momentum reflects the fact that the scattering amplitude
starts at order $\alpha_s^2$ in perturbation theory, hence one needs some
non--trivial rapidity evolution $Y_0$ before the dipole becomes `black'
on the resolution scale fixed by its own size: $T_{\rm
dd}(r=R,R,Y_0)=\kappa$. This condition yields\footnote{In the
Introduction, this was written as $Y_0\simeq
(1/\omega_\mathbb{P})\ln(1/\alpha_s^2)$, which is essentially the same,
since $\gamma_s\lambda_s\abar$ plays the role of the intercept at
the saturation saddle point, as manifest in Eq.~\eqref{TSAT}.} $Y_0\simeq
(1/\gamma_s\lambda_s\abar)\ln(\kappa/\alpha_s^2)$, and then
Eq.~\eqref{QSAT} can be rewritten in such a way to exhibit the rapidity
excess beyond $Y_0$:
 \beq\label{QSAT1}
  Q_s^2(R,Y)\,=
  \,\frac{\rme^{\lambda_s\abar (Y-Y_0)}}{R^2}\,.
  \eeq
Eq.~\eqref{TSAT} is valid when $\abar Y\gg 1$ and in a rather wide range
of values for $r$, namely for $1 \ll \ln(1/r^2 Q_s^2)\lesssim c\abar Y$,
with $c\sim\order{1}$. In particular, within the more restricted window
 \beq\label{gswindow}
 1\,\ll\, \ln\frac{1}{r^2 Q_s^2}\,\ll\,\sqrt{4D_s\abar Y}\,,\eeq
the last, Gaussian, factor in Eq.~\eqref{TSAT} can be ignored, and then
the amplitude shows geometric scaling, as anticipated. Note that, when
$\abar Y\gg 1$, this scaling window is quite wide, especially since the
parameter $D_s$ is numerically large.

\section{Fixed coupling case: Mueller--Navelet jets}
\setcounter{equation}{0} \label{Sect:FCMN}

We now have all the ingredients to compute the cross--section for
Mueller--Navelet jets in the presence of unitarity corrections and for
fixed coupling. We shall focus on the range of values for the momentum
cutoffs $Q_1$ and $Q_2$ in which we expect geometric scaling. Namely we
choose $Q_1\gg Q_2$ in such a way that $Q_1^2\gtrsim Q_2^2\,
\rme^{\lambda y}$, with the compact notations
$\lambda\equiv\lambda_s\abar$ and $y\equiv Y-Y_0$. That is, $y$ is the
rapidity excess introduced in Eq.~\eqref{QSAT1}. (Throughout this paper,
we assume $Y>Y_0$.). Therefore, $Q_1$ is larger, but not much larger,
than the saturation scale that would generated by a target dipole with
size $\sim 1/Q_2$ after a rapidity evolution $Y$. As we shall shortly
check, under this condition the convolutions in Eq.~\eqref{j1j1} are
indeed dominated by values for $r_1$ and $r_2$ within the scaling window
(\ref{gswindow}).

To simplify the calculation, we therefore keep only the scaling piece in
the dipole--dipole amplitude \eqref{TSAT} in the weak scattering regime.
We therefore replace Eqs.~\eqref{Tgg}--\eqref{TSAT} by the following,
piecewise, approximation to the dipole--dipole cross--section (we recall
the notations $r \equiv\min(r_1,r_2)$ and $R \equiv\max(r_1,r_2)$)
    \beq\label{ssat}
    \sigma_{\rm dd}(r_1,r_2,Y) = 2\pi R^2\,
    \begin{cases}
    \left(\displaystyle{\frac{r^2}{R^2}\ \rme^{\lambda y}}
    \right)^{\gamma_s} \quad &\text{for}
    \quad r^2 < R^2\, \rme^{-\lambda y} \\*[0.5cm]
    1 \quad &\text{for} \quad r^2 > R^2\, \rme^{-\lambda y},
    \end{cases}
    \eeq
where we have ignored the slowly varying logarithm in Eq.~\eqref{TSAT}.
The precise normalization of the cross--section is not an issue here, as
we are merely interested in its functional dependencies.

Because of the symmetry of the above cross--section under $r_1
\leftrightarrow r_2$, when evaluating the double integral in
Eq.~\eqref{j1j1} it is enough to consider the case $r_1 < r_2$. Then the
contribution from the other region $r_1 > r_2$ can be simply obtained by
letting $Q_1 \leftrightarrow Q_2$ in the result of the first
case\footnote{Given our choice that $Q_1\gg Q_2$, one could anticipate
that the dominant contribution comes from the region $r_1\ll r_2$. For
completeness, we shall nevertheless consider the region $r_1> r_2$ too.}.
Thus, substitution of \eqref{ssat} into \eqref{j1j1} gives
    \beq\label{MNJgsa}
    \frac{\dif \sigma^{pp \rightarrow J X J}}{\dif x_1 \dif x_2} = & &
    F\, \rme^{\gamma_s \lambda y}
    \int\limits_0^{\infty}
    \dif r_1\, Q_1 J_1(Q_1 r_1) r_1^{2 \gamma_s}\!\!\!\!
    \int\limits_{r_1\rme^{\lambda y/2}}^{\infty}\!\!\!\!
    \dif r_2 \,Q_2 J_1(Q_2 r_2) r_2^{2-2 \gamma_s}
    \nn
    & & +\, F
    \int\limits_0^{\infty}
    \dif r_1\, Q_1 J_1(Q_1 r_1)\!\!\!\!
    \int\limits_{r_1}^{r_1\rme^{\lambda y/2}}\!\!\!\!
    \dif r_2 \,Q_2 J_1(Q_2 r_2) r_2^2\, +\, \{Q_1 \leftrightarrow Q_2 \},
    \eeq
where for the time being we consider the gluon jets alone (hence, the
overall factor $F$ includes the gluon distributions, together with other
numerical factors); the quark jets will be added later on. Note that the
exchange $Q_1 \leftrightarrow Q_2$ should not be done inside $F$, but
only in the result of the integration. Clearly, the two explicit terms in
the r.h.s. of the above equation arise from the corresponding pieces in
\eqref{ssat}. It is convenient to change variables by letting $u_1 = Q_1
r_1$ and $u_2 = Q_2 r_2$. Then the above equation becomes
    \beq\label{MNJgsb}
    \frac{\dif \sigma^{pp \rightarrow J X J}}{\dif x_1 \dif x_2} = & &
    F\,\frac{1}{Q_2^2} \left(\frac{Q_2^2\, \rme^{\lambda y}}{Q_1^2} \right)^{\gamma_s}
    \int\limits_0^{\infty}
    \dif u_1 \,u_1^{2\gamma_s} J_1(u_1)\!
    \int\limits_{a u_1}^{\infty}\!
    \dif u_2\, u_2^{2-2 \gamma_s} J_1(u_2)
    \nn
    & & +\, F\,
    \frac{1}{Q_2^2}
    \int\limits_0^{\infty}
    \dif u_1\, J_1(u_1)\!
    \int\limits_{b u_1}^{a u_1}\!
    \dif u_2\, u_2^2 J_1(u_2) \, +\, \{Q_1 \leftrightarrow Q_2 \},
    \eeq
where we set $a = (Q_2/Q_1)\exp(\lambda y /2)$ and $b=Q_2/Q_1$. In what
follows we shall show that the only non--zero term in the r.h.s. of
Eq.~\eqref{MNJgsb} is the first one, and moreover this term shows
geometric scaling.

The fact that the second term in \eqref{MNJgsb} vanishes could have been
anticipated, since this term arises from the saturation piece in
Eq.~\eqref{ssat}, which in turn depends only on one of the two variables
$r_1$ and $r_2$. Hence, if we return to the unintegrated version of the
cross--section, Eq.~\eqref{MVjets}, it becomes obvious that this
saturation piece goes away by the successive action of the two Laplacians
$\grad^2_{\bm{r}_1}\grad^2_{\bm{r}_2}$. It is a little bit more difficult
to see the corresponding cancelation in \eqref{MNJgsb}. Performing the
integration over $u_2$ we obtain for this second term under consideration
    \beq\label{second}
    F\,\frac{1}{Q_2^2}
    \int\limits_0^{\infty}
    \dif u_1\,u_1^2 J_1(u_1) \left[a^2 J_2(a u_1) - b^2 J_2(b u_1)\right].
    \eeq
To evaluate this integral, let us choose $n=1$ in the completeness
formula
    \beq\label{completeness}
    \int\limits_{0}^{\infty} \dif u \, u J_n(u) J_n(a u ) = \delta(a-1),
    \eeq
then differentiate this identity with respect to $a$ and use the fact
that $J_1'(x) = J_1(x)/x - J_2(x)$. We thus obtain
    \beq\label{Bessel1}
    \int\limits_{0}^{\infty} \dif u \, u^2 J_1(u) J_2(a u ) =
    \delta(a-1) - \delta'(a-1),
    \eeq
which provides the result for the integrals in Eq.~\eqref{second}.
Namely, since we are interested in momenta such that $Q_1^2 > Q_2^2
\exp(\lambda y)$, we see that $a<1$ and $b<1$, hence both terms in
Eq.~\eqref{second} vanish, as anticipated.

Let us now turn to the calculation of the first term in \eqref{MNJgsb}.
Putting aside the prefactor, the remaining double integration, let us
call it $h(a)$, can be simplified by differentiating with respect to $a$.
We have
    \beq\label{hprimea}
    h'(a) = - a^{2- 2 \gamma_s}
    \int\limits_{0}^{\infty} \dif u \, u^3 J_1(u) J_1(a u ).
    \eeq
The above integral can be again recognized as the derivative of a known
integral: by differentiating \eqref{Bessel1} with respect to $a$ and
using $J_2'(x) = J_1(x) - 2 J_2(x)/x$, we find
    \beq\label{Bessel2}
    \int\limits_{0}^{\infty} \dif u \, u^3 J_1(u) J_1(a u ) =
    - \delta'(a-1) - \delta''(a-1).
    \eeq
It is now straightforward to obtain $h(a)$ by integrating over $a$.
Integrating by parts to get rid of the derivatives acting on the
$\delta$--function and using $h(\infty) = 0$, we obtain
    \beq\label{ha}
    h(a) = 4 \gamma_s (1- \gamma_s) \Theta(1 - a).
    \eeq
Recalling that $a<1$ in the kinematical region of interest, we see that
the step function is equal to 1. That is, the first (double) integral in
the r.h.s. of \eqref{MNJgsb} is independent of $a$ so long as $a<1$.

By a similar argument, it is now easy to see that the term obtained by
exchanging $Q_1 \leftrightarrow Q_2$ (that is, the term coming from the
region $r_1>r_2$) is equal to zero: indeed, the corresponding
contribution would be proportional to $\Theta(1 - 1/a)$.

Now, whereas the above, exact, results are of course attributed to the
precise form of the interpolation chosen in Eq.~\eqref{ssat} for the
dipole--dipole cross--section, it is clear that these results will
approximately hold for {\em any} smooth interpolation (in between the
shown limiting expressions) provided we impose the {\em strong}
inequality $Q_1^2 \gg Q_2^2 \exp(\lambda y)$. Therefore, putting
everything together, we arrive at (recall that $y=Y-Y_0$)
    \beq\label{MNJsat}
    \frac{\dif \sigma^{pp \rightarrow J X J}}{\dif x_1 \dif x_2}
    \,\simeq\,
    F_{\rm eff}\,\frac{1}{Q_2^2} \left(\frac{Q_2^2\, \rme^{\lambda y}}{Q_1^2}
    \right)^{\gamma_s}
    \qquad \text{for} \qquad
    Q_1^2\, \gg\, Q_2^2 \,\exp(\lambda y),
    \eeq
valid up to an overall, numerical, factor which is not under control. The
effective parton distribution $F_{\rm eff}$ of Eq.~\eqref{Feff} has been
generated because the saturation momenta are, strictly speaking,
different for different types of dipoles, as explained below
Eq.~\eqref{QSAT}, and these differences can be absorbed in the
normalization of the parton distributions, as shown in Eq.~\eqref{Feff}.

Apart the prefactor $F_{\rm eff}$, the above expression can be obtained
from the weak--scattering piece in Eq.~\eqref{ssat} via the replacements
$r \to 1/Q_1$ and $R \to 1/Q_2$. The dimensionfull factor $1/Q_2^2$ plays
the role of the `area of the larger (target) dipole', whereas the
dimensionless ratio ${Q_2^2\, \rme^{\lambda y}}/{Q_1^2}$ is recognized as
the scaling variable $\tau\equiv Q_s^2(Y)/{Q_1^2}$, with $Q_s(Y)$ the
saturation momentum of this `target dipole'. Accordingly,
Eq.~\eqref{MNJsat} exhibits {\em geometric scaling}, as anticipated: it
depends upon the resolution $Q_1^2$ of the `small dipole' and the
rapidity $y$ only via the scaling variable $\tau$. The validity region
for this behavior should be clear too from the previous manipulations:
since the effect of the Fourier transforms is to select $r_1 \sim 1/Q_1$
and $r_2 \sim 1/Q_2$ (at least, so long as the external momenta $Q_1$ and
$Q_2$ are well separated from each other), it is quite clear that the
geometric scaling in the cross--section for Mueller--Navelet jets at
fixed coupling holds in the same kinematical window as for the
dipole--dipole scattering amplitude, that is,
 \beq\label{MNwindow}
 1\,\ll\, \ln\frac{Q_1^2}{Q_s^2(Q_2,y)}\,\ll\,\sqrt{4D_s\abar Y}\,.\eeq
By the same argument, we also expect geometric scaling behavior in the
`unintegrated' cross--section \eqref{MVjets}, for transverse momenta
$k_{1\perp}$ and $k_{2\perp}$ replacing $Q_1$ and $Q_2$ in the above
formul\ae.

\section{Running coupling}
\setcounter{equation}{0}

With a running coupling, the theoretical situation is less firmly under
control, since the NLO formalism is not yet fully developed for the
unitarity corrections. (The running--coupling version of the BK equation
became available only recently \cite{KWrun1,Brun,BalitChir07}.) Still,
for the specific problem at hand, we need only some limited information
about the NLO effects, that we believe to be reliably described by the
present formalism. Indeed, to study geometric scaling in the
Mueller--Navelet jets, we need the dipole--dipole cross--section in the
weak scattering regime, where the BFKL approximation is expected to
apply, and for which the NLO formalism is by now well established
\cite{NLBFKL,Salam99} (including the approach towards saturation
\cite{MT02,DT02}). As mentioned in the Introduction, the use of the BFKL
approximation is better justified in the running--coupling scenario than
in the fixed--coupling one, since the effects of gluon--number
fluctuations (which tend to invalidate this approximation) are
drastically suppressed by the running of the coupling \cite{RPLOOP}. In
fact, as we shall shortly discover, the main obstruction to our
calculation does not come from our limited knowledge of the NLO
perturbative formalism, but rather from a drawback of perturbation theory
itself, ultimately associated with the running of the coupling. To
identify this difficulty, we start by assuming perturbation theory to
apply.

As before, we choose hard cut--off momenta, $Q_1,Q_2\gg\Lam$, with
moreover $Q_1\gg Q_2$. In the fixed--coupling case, this condition was
enough to ensure that the relevant dipole sizes are sufficiently small,
$r_i\sim 1/Q_i\ll 1/\Lam$, for perturbation theory to apply. With a
running coupling, this strong correlation between $Q_i$ and $r_i$ is lost, as we shall see, but for the time being let us simply assume that the
(effective) dipoles are perturbatively small. The typical situation is
such that one dipole is much larger than the other, $R\gg r$, and we
shall assume, once again, that the larger dipole (the `target') evolves
towards high gluon density and blackness on transverse sizes $r\le R$
much faster than it expands in impact parameter space. Under these
assumptions, the dipole--dipole cross--section is again given by
Eq.~\eqref{Tgg}, but with the scattering amplitude $T_{\rm dd}(r,R,Y)$
now computed for a running coupling. This calculation has been described
somewhere else \cite{MT02,DT02,IIT04}, and here we present only the
relevant results.

\texttt{(i)} The saturation momentum is now estimated as
   \beq\label{rsat}
    Q_s^2(R,Y) = \Lam^{2} \exp\left[{\sqrt{2 c (Y-Y_0)
    + \rho_R^2}}\,\right]\,,\qquad\rho_R\,\equiv\,\ln\frac{1}{R^2 \Lam^2}
    \eeq
where the QCD scale $\Lam$ has been introduced via the running of the
coupling, for which we used the one--loop result
$\alpha_s(Q^2)=b_0/\ln(Q^2/\Lam^2)$. We have denoted $c\equiv
b_0N_c\lambda_s/\pi$, with the saturation exponent $\lambda_s\approx
4.88$ (the same as in Eq.~(\ref{QSAT})). As before, $Y_0$ denotes the
rapidity evolution necessary to build a saturation scale $Q_s$ equal to
$1/R$.

More precisely, Eq.~(\ref{rsat}) has been obtained by interpolating
between the asymptotic behavior at large $Y$, where the calculation is
better under control, and the expected behavior at low $Y$, where one
should recover the fixed--coupling result (\ref{QSAT1}). Indeed, for $2 c
y\ll \rho_R^2$, with $y\equiv Y-Y_0$, Eq.~(\ref{rsat}) reduces to
$Q_s^2\simeq (1/R^2) \exp(\lambda_s\abar y)$, with $\abar$ evaluated at
$Q^2=1/R^2$. On the other hand, for energies high enough such that $2 c
y\gg \rho_R^2$, the saturation momentum loses any dependence upon the
target size \cite{AM03}:
\beq\label{rsatas}
    Q_s^2(R,Y) \,\simeq\, Q_c^2(Y) \,\equiv\,\Lam^{2} \,\rme^{\sqrt{2 c y
    }}\ \quad\mbox{when}\ \quad2 c y\gg \rho_R^2\,.
    \eeq

\texttt{(ii)} Consider the scattering amplitude $T_{\rm dd}(r,R,Y)$ for a
projectile dipole with size $r$. Within the relatively wide region at
 \beq\label{BFKLrun}
 1\,\ll\, \ln\frac{1}{r^2 Q_s^2}\,\lesssim \,(2 c y
    + \rho_R^2)^{1/3}\,,\eeq
where the scattering is weak ($T_{\rm dd}\ll 1$), this amplitude is a
universal function of the `scaling' variable $\tau\equiv r^2Q_s^2(R,Y)$
and of $Y$, whose structure is quite similar\footnote{See Eq.~(3.27) in
Ref. \cite{IIT04} for the specific function in the case of a running
coupling.} to that at fixed coupling, cf. Eq.~\eqref{TSAT}: namely, it
involves the power $\tau^{\gamma_s}$ times a function of $\tau$ and $Y$
which violates geometric scaling via a diffusive pattern.

\texttt{(iii)}  Within the more restricted window at
\beq\label{gswindowrun}
 1\,\ll\, \ln\frac{1}{r^2 Q_s^2}\,\lesssim\, (2 c y
    + \rho_R^2)^{1/6}\,,\eeq
the scaling violations can be neglected, and the amplitude takes the same
scaling form as in the fixed--coupling case, that is,
 \beq\label{Tscal}
 T_{\rm dd}(r,R,Y)\,\simeq\,\big[r^2 Q_s^2(R,Y)\big]^{\gamma_s}\,.\eeq

Two important observations about the above results are here in order:
\texttt{(i)} The running of the coupling considerably slows down the
evolution, as clear from the fact that both the saturation momentum
\eqref{rsat} and the width of the scaling region \eqref{gswindowrun} rise
much slower with $Y$ than at fixed coupling. \texttt{(ii)} In the
high--energy regime where Eq.~\eqref{rsatas} applies, the dipole--dipole
amplitude is insensitive to the target dipole size within the whole
validity range for the BFKL approximation (cf. Eq.~\eqref{BFKLrun}). This
second observation has dramatic consequences for the Mueller--Navelet
process, to which we now return.

For simplicity, we keep only the scaling piece \eqref{Tscal} in the
dipole--dipole amplitude (as we shall later argue, our main results are
independent of this approximation), and thus write (compare to
Eq.~\eqref{ssat} at fixed coupling)
    \beq\label{srun}
    \sigma_{\rm dd}(r_1,r_2,Y) \,\approx\,2\pi R^2\,
    \begin{cases}
    \left(\displaystyle{\frac{r^2}{r_s^2}}\right)^{\gamma_s}
    \quad &\text{for} \quad r < r_s \\*[0.5cm]
    1 \quad &\text{for} \quad r > r_s.
    \end{cases}
    \eeq
where $r_s\equiv 1/Q_s(R,Y)$ encodes the whole dependence upon both the
rapidity and the target size. Below, we shall also use $r_c\equiv
1/Q_c(Y)$, cf. Eq.~\eqref{rsatas}.

Let us calculate the contribution of the weak--scattering piece in
Eq.~\eqref{srun} (the first line there) to the Mueller--Navelet dijet
cross section. (As in the fixed--coupling case, one can show that the
respective contribution coming from the saturation piece is equal to
zero.) We focus on the case $r_2 > r_1$, which we expect to be the most
interesting one, in view of our condition that $Q_1\gg Q_2$. Hence, from
now on, $r=r_1$ and $R=r_2$, and we have
    \beq\label{MNJruna}
    \frac{\dif \sigma^{pp \rightarrow J X J}}{\dif x_1 \dif x_2} =
    F
    \int\limits_{r_1<r_s}\!\!\!
    \dif r_1\dif r_2\, Q_1 J_1(Q_1 r_1) \,Q_2 J_1(Q_2 r_2)\,
    r_2^2\,\left(\frac{r_1^2}{r_s^2}\right)^{\gamma_s}.
    \eeq
It is convenient to distinguish between two regions of integration over
$r_2$: \texttt{(a)} $r_2>r_c$ and \texttt{(b)} $r_2<r_c$. Given that
$r_c^2\propto \exp\left(-\sqrt{2c y}\right)$ is rapidly decreasing with
$y$, it is quite clear (and easy to check) that the parametrically
dominant contribution at large $y$ is the one coming from region
\texttt{(a)}. In this region, we typically have $r_2\gg r_c$, so that the
saturation scale $r_s\approx r_c$ is independent of the size $r_2$ of the
target. (Indeed, the condition $r_2\gg r_c$ is the same as $\rho_R^2\ll
2cy$, cf. Eq.~\eqref{rsatas}, applied to $R=r_2$.) Then
Eq.~\eqref{MNJruna} becomes
    \beq\label{MNJrunb}
    \frac{\dif \sigma^{pp \rightarrow J X J}}{\dif x_1 \dif x_2} \approx
    F\,\frac{1}{r_c^{2\gamma_s}}\,
    \int\limits_{0}^{r_c}
    \dif r_1\, Q_1 J_1(Q_1 r_1) r_1^{2\gamma_s}
    \int\limits_{r_c}^{\infty}
    \dif r_2\, Q_2 J_1(Q_2 r_2)
    r_2^2 \,\, +\,\, \dots,
    \eeq
with the dots standing for the contributions coming from dipoles of size
$r_2 \lesssim r_c$. We do our standard change of variables $u_1 = Q_1
r_1$ and $u_2 = Q_2 r_2$ to obtain
    \beq\label{MNJrunc}
    \frac{\dif \sigma^{pp \rightarrow J X J}}{\dif x_1 \dif x_2} \approx
    F\,\frac{1}{Q_2^2}\,\left(\frac{Q_c^2}{Q_1^2}\right)^{\gamma_s}
    \int\limits_{0}^{Q_1/Q_c}\!\!\!
    \dif u_1\, u_1^{2\gamma_s} J_1(u_1) \!\!\!
    \int\limits_{Q_2/Q_c}^{\infty}\!\!\!
    \dif u_2\,u_2^2 J_1(u_2)
     \,\, +\,\, \dots,
    \eeq
Naturally, the regime that we are interested in is $Q_1 \gg Q_c(Y) \gg
Q_2$. In this regime, $Q_c(Y)$ is essentially the same as the saturation
scale for a target dipole with size $1/Q_2$. Therefore, the prefactor
appearing outside the integrations in Eq.~\eqref{MNJrunc} has the right
structure to exhibit geometric scaling. There are, of course, additional
functional dependencies in the limits of the remaining integrations, but
at a first sight it seems that these dependencies are rather weak and
therefore negligible: Since $Q_1 \gg Q_c$, we can extend the upper limit
of the $u_1$ integration to $\infty$; then, by also making use of
\eqref{jpower}, one sees that this integration yields a positive number
of $\order{1}$. Similarly, since $Q_2 \ll Q_c$, we can extend the lower
limit of the $u_2$ integration to 0. But the problem that we are facing
then is that, according to Eq.~\eqref{jpower}, the result of the ensuing
integration over $u_2$ is exactly zero.

Thus, by making approximations aiming at preserving the dominant
contributions to the Mueller--Navelet cross--section at running coupling,
we have found a result which is identically zero. Of course, a non--zero
result could be instead obtained by keeping the formerly discarded (since
formally subleading) contributions. But would that result be correct
indeed? We do not believe so since, first, that result would be
generated by physically implausible corners of the phase--space and,
second, it would be strongly sensitive to the fine details of our
approximations --- it could even oscillate between positive and negative
values, an unacceptable feature  for a cross--section.

This invites us to critically reexamine the above calculation, in order
to better understand why we obtained this vanishing result. At a
mathematical level, this is related to the oscillatory behavior of the
Bessel functions. The relevant integral, that is $\int_0^\infty \dif
u_2\,u_2^2 \,J_1(u_2)$, is quite peculiar: for large values of the
variable $u_2$, the oscillations of $J_1(u_2)$ are strongly amplified by
the factor $u_2^2$. Hence, if the overall result turns out to be zero, it
is because of exact cancelations between large contributions with
opposite signs. Then, clearly, the result of this integration is
controlled by the behavior of its integrand at large $u_2\gg 1$. If one
sharply cuts off the integral at some value $u_2^{\rm max}\gg 1$, then
the result is an oscillating function of $u_2^{\rm max}$, which can vary
from large positive values to large negative ones.

We see that, even though we have tried to set up a perturbative
calculation, the final, vanishing, result that we have obtained is in
fact controlled by large dipoles --- in fact, {\em arbitrarily} large
---, for which the perturbative approach is not justified anymore. At
this point, one may wonder about the real significance of this result:
does it signal a true failure of perturbation theory, or is this merely
an artifact of our specific approximations? The following argument,
based on a comparison with the situation at fixed coupling, suggests that
the first answer should be the correct one.

The corresponding integral in the fixed--coupling case, namely
$\int_0^{\infty} \dif u_2\, u_2^{2-2 \gamma_s} J_1(u_2)$ (see
Eq.~\eqref{MNJgsb}), was convergent and dominated by $u_2\sim 1$ (i.e.,
$r_2\sim 1/Q_2$) because, in the dipole--dipole cross--section
\eqref{Tgg}, the rapid growth $\propto r_2^2$ of the target area was
partially compensated by the decay $\propto (1/r_2)^{2 \gamma_s}$ of the
scattering amplitude $T_{\rm dd}$ at very large $r_2$ (cf.
Eqs.~(\ref{TSAT})--(\ref{QSAT}) with $r\to r_1$ and $R\to r_2$). In that
case, the dipole--dipole amplitude {\em at a given impact parameter} can
be made arbitrarily small be increasing the overall size of the target.
This is a very peculiar feature of the leading--order formalism,
ultimately related to its conformal invariance: the target size is the
only dimensionfull parameter in the problem, so the gluon density in the
target, as measured by the (local) saturation momentum \eqref{QSAT}, must
be proportional to an appropriate power of $1/r_2$.

The situation changes at NLO, where the running of the coupling
introduces an additional mass scale in the problem, the `soft' scale
$\Lam$. Then, for sufficiently high energy, the local saturation momentum
becomes insensitive to the overall target size, as manifest on
Eqs.~\eqref{rsat}--\eqref{rsatas}. This result is quite natural: the
local gluon distribution is determined by the physics on the distance
scale $1/Q_s(Y)$, which decreases with increasing $Y$,
and is in any case much smaller than the target size $r_2$. Similarly,
the amplitude $T_{\rm dd}(r_1,r_2,Y)$ for a small dipole ($r_1
< 1/Q_s \ll r_2$) and for large enough $Y$ is insensitive to $r_2$, as
emphasized after Eq.~\eqref{Tscal}. This property holds within the
validity range \eqref{BFKLrun} of the BFKL approximation, and not only
within the narrower window \eqref{gswindowrun} for geometric scaling.
Accordingly, the above conclusion about Eq.~\eqref{MNJrunc} is more general than the geometric--scaling ansatz in Eq.~\eqref{srun}:
within the whole range in which the BFKL approximation is expected to be
valid, the (perturbative) dipole--dipole cross--section for two dipoles
with very disparate sizes is expected to grow like the area $\sim r_2^2$
of the larger dipole. The growth is so fast that the cross--section for
Mueller--Navelet jets is ineluctably dominated by the largest possible
`target' dipoles, whose treatment goes beyond the scope of perturbation
theory.

Of course, in QCD dipoles cannot become arbitrarily large,
because of confinement. In what follows we propose a heuristic
modification of the previous calculation which limits the dipole sizes to
a value $\sim 1/\Lam$ and thus yields a finite result for the
Mueller--Navelet dijet cross--section. Clearly, the precise value of this
result will be sensitive to our specific prescription for introducing
confinement, and we shall try to  motivate this
prescription on physical grounds. But before we proceed, it is important
to emphasize that this prescription will affect only the
$u_2$--integration in Eq.~\eqref{MNJrunc}, but not also the functional
dependencies upon $Q_1$ and $Y$, as encoded in the prefactor there. In
other terms, whatever prescription we choose to eliminate the large
target dipoles, this will not change the geometric scaling behavior of
the cross--section, as determined by the prefactor.

To motivate our prescription for introducing confinement, let us
recall that the `dipoles' under consideration are {\em effective}
dipoles, built with one gluon at $\bm{x}$ in the amplitude and
another gluon at $\bm{y}$ in the complex conjugate amplitude, and such
that $R=|\bm{x}-\bm{y}|$. Hence, the maximal possible value for $R$ is
the same as the maximal dispersion between the positions in impact
parameter space at which a gluon can be produced in the proton
wavefunction. This distance is of the order of the proton size $\sim
1/\Lam$. Moreover, larger impact parameters $\gtrsim 1/\Lam$ lie in the
tail of the proton wavefunction, where the gluon distribution must decay
exponentially, so as it happens for any quantum--mechanical system with a
mass gap. Thus we conjecture that the probability to produce an effective
dipole with large size $R\gtrsim 1/\Lam$ should fall exponentially,
according to $\exp(-R\Lam)$. We can implement this prescription in our
calculation via the following replacement for the area factor in
Eq.~\eqref{srun} (with $\Lambda\equiv\Lam$ from now on):
    \beq\label{conf}
     R^2 \to R^2\exp(-R\Lambda).
    \eeq
Then it is straightforward to see that the last integration in
\eqref{MNJrunc} becomes
    \beq\label{confint}
    \int\limits_{Q_2/Q_c}^{\infty}\!\!\!
    \dif u_2\,u_2^2 J_1(u_2) \exp(-u_2 \Lambda/Q_2) \,\simeq\,
    \frac{3 \Lambda}{Q_2}\,\frac{1}
    {(1+\Lambda^2/Q_2^2)^{5/2}} - \frac{1}{8}
     \frac{Q_2^4}{Q_c^4} \,\simeq \,\frac{3 \Lambda}{Q_2},
    \eeq
where the approximate equality holds when $\Lambda/Q_2\gg
{Q_2^4}/{Q_c^4(Y)}$ (with $\Lambda/Q_2\ll 1$, though), a situation which
is eventually reached with increasing energy at fixed $Q_2$. Under these
assumptions, the high--energy behavior of the Mueller--Navelet
cross--section reads
    \beq\label{MNJconf}
    \frac{\dif \sigma^{pp \rightarrow J X J}}{\dif x_1 \dif x_2}
    \,\simeq \,F_{\rm eff}\,
    \frac{\Lambda}{Q_2^3}\,
    \left(\frac{Q_c^2(Y)}{Q_1^2}\right)^{\gamma_s}\,,
    \eeq
a formula which should be valid for sufficiently large $Y$ and for $Q_1
\gg Q_c \gg Q_2 \gg \Lambda$. The overall normalization factor
${\Lambda}/{Q_2^3}$ in this equation depends, of course, upon our
specific model for introducing confinement, but the scaling behavior
w.r.t. $\tau\equiv {Q_c^2(Y)}/{Q_1^2}$ does not. We therefore consider
this scaling behavior as a robust prediction of our analysis. More
precisely, this behavior should hold within the window
  \beq\label{GSWINDOW}
 1\,\ll\, \ln\frac{Q_1^2}{Q_c^2(Y)}\,\lesssim\, (2 c y)^{1/6}\,,
 \eeq
which is obtained after replacing $r\equiv r_1\to 1/Q_1$ and $Q_s\to Q_c$
in Eq.~\eqref{gswindowrun}. (Such a replacement is legitimate, since the
integration over $r_1$ in Eq.~\eqref{MNJrunb} is indeed dominated by
$r_1\sim 1/Q_1$.)

\section{Conclusion and perspectives}
\setcounter{equation}{0}

Our main result in this paper is that, under specific kinematical
conditions --- namely, for a sufficiently large rapidity gap $Y$ and for
a sufficiently pronounced asymmetry between the transverse momenta of the
two jets --- the partonic core of the cross--section for Mueller--Navelet
jets should exhibit geometric scaling.

In the leading--order formalism, where the coupling is fixed, this result
is a rather straightforward consequence of the factorization
(\ref{MVjets}) for the dijet cross--section together with known results
about the dipole--dipole scattering within the framework of the BK (or
Balitsky--JIMWLK) equation. Although the factorization (\ref{MVjets}) has
not been established here in full rigor, this should not affect the
generality of our analysis, which employed Eq.~(\ref{MVjets}) only in the
weak scattering regime where the $k_T$--factorization is firmly
established (at LO). However, as explained in the Introduction, this
fixed--coupling analysis is a bit academic since, first, within a
complete LO calculation its conclusions would be affected by
particle--number fluctuations (at least, for sufficiently high energy)
and, second, in real QCD the coupling is running anyway.

With a running coupling, on the other hand, our analysis lacks rigor at
several points --- it neglects other NLO corrections except for the
running of the coupling and, especially, it turns out to transcend the
framework of perturbation theory --- and hence should be viewed as merely
exploratory. Yet, we believe that our main conclusion (concerning the
emergence of geometric scaling) is rather robust even in that context,
because it is mainly based on the analysis of the harder jet, for which
perturbation theory appears to be reliable. As explained in Sect. 5, the
failure of perturbation theory refers merely to the softer jet, and, more
precisely, to the connection between the transverse momentum $Q_2$ of
that jet and the size $r_2$ of the associated, effective, dipole: even
when $Q_2$ is relatively hard, $Q_2^2\gg \Lam^2$, the cross--section for
dijet production is still dominated by very large `target' dipoles, with
$r_2\sim 1/\Lam$, because the perturbative dipole--dipole cross--section
grows very fast with $r_2$. This growth is faster with a running coupling
since the corresponding saturation momentum is independent of the target
size.

This failure of perturbation theory is perhaps a bit surprising, as this
is not the usual failure associated with the BFKL `infrared diffusion' in
the presence of a running coupling: as expected, gluon saturation
eliminates the IR diffusion and sets the argument of the coupling to a
relatively hard scale $\sim Q_s(Y)$, so that the Landau pole in the
coupling is not an issue any more. In spite of that, an infrared problem
remains, as alluded to above, and its identification can be viewed as our
second main result.

Since based on asymptotic expansions, our results can be trusted,
strictly speaking, only for sufficiently high energies, so their
applicability to LHC may be questionable. This being said, and in view of
the rather successful phenomenology at HERA, it would be nevertheless
interesting to look for traces of this geometric scaling behavior in the
forthcoming data at LHC. For instance, while keeping fixed the kinematics
$(Q_2,\eta_2)$ of the softest jet, one could vary the the transverse
momentum cutoff $Q_1$ (with $Q_1>Q_2$) and the pseudo--rapidity $\eta_1$
of the hardest jet, in such a way to preserve a constant value for the
respective longitudinal momentum fraction $x_1\simeq
({Q_{1}}/{\sqrt{s}})\rme^{\eta_1}$. In this way, $Q_1$ and the rapidity
gap $Y$ would be simultaneously changing, and then one could check
whether the ensuing variation in the measured dijet cross--section
follows indeed a geometric scaling pattern, cf. Eq.~\eqref{MNJconf}, at
least approximately. Of course, such a scaling should be partially
violated by the $Q_1$--dependence of the parton distributions within
$F_{\rm eff}$, but this dependence should be rather weak and, in any
case, controllable within perturbation theory.

\section*{Acknowledgments}

We would like to thank Al Mueller for fruitful discussions and insightful
remarks.


\end{document}